# Nanocrystalline equiatomic CoCrFeNi alloy thin films: Are they single phase fcc?

Maya K. Kini[1*], Subin Lee[1], Alan Savan[2], Benjamin Breitbach[1], Younes Addab[3,4], Wenjun Lu[1], Matteo Ghidelli[1,5], Alfred Ludwig[2], Nathalie Bozzolo[4], Christina Scheu[1], Dominique Chatain[3], Gerhard Dehm[1*]

1 Max-Planck-Institut für Eisenforschung GmbH, Düsseldorf, Germany

2 Materials Discovery and Interfaces, Institut für Werkstoffe, Ruhr-Universität Bochum, Germany

3 Aix-Marseille Univ, CNRS, CINAM, 13009 Marseille, France

4 Mines ParisTech PSL-Research University, Centre de Mise en Forme des Matériaux (CEMEF), CNRS UMR 7635, CS 10207, 06904, Sophia Antipolis Cedex, France

5 Current Address: Laboratoire des Sciences des Procédés et des Matériaux (LSPM), CNRS, Université Sorbonne Paris Nord, 93430, Villetaneuse, France

*Corresponding authors

Email id: m.kini@mpie.de , g.dehm@mpie.de

**Abstract**

The bulk quaternary equiatomic CoCrFeNi alloy is studied extensively in literature. Under experimental conditions, it shows a single-phase fcc structure and its physical and mechanical properties are similar to those of the quinary equiatomic CoCrFeMnNi alloy. Many studies in literature have focused on the mechanical properties of bulk nanocrystalline high entropy alloys or compositionally complex alloys, and their microstructure evolution upon annealing. The thin film processing route offers an excellent alternative to form nanocrystalline alloys. Due to the high nucleation rate and high density of defects in thin films synthesized by sputtering, the kinetics of microstructure evolution is often accelerated compared to those taking place in the bulk. Here, thin films are used to study the phase evolution in nanocrystalline CoCrFeNi deposited on Si/SiO$_2$ and c-sapphire substrates by magnetron co-sputtering from elemental sources. The phases and microstructure of the films are discussed in comparison to the bulk alloy. The main conclusion is that second phases can form even at room temperature provided there are sufficient nucleation sites.

## 1. Introduction

The field of high entropy alloys (HEAs) or compositionally complex alloys (CCAs) has received great interest since the first report by Cantor *et al.* [1]. Although a variety of elemental combinations are investigated in literature [2, 3], many studies are based on varying compositions and processing conditions around the quinary equiatomic alloy of CoCrFeMnNi. The quinary equiatomic CoCrFeMnNi 'Cantor' alloy (referred to as 'quinary alloy' hereafter), was found to form a single-phase fcc solid solution in the original work [1]. In a few follow up studies a quaternary equiatomic CoCrFeNi alloy (referred to as 'quaternary alloy' hereafter) was also found to be a single-phase fcc solid solution, with most physical and mechanical properties similar to those of the quinary alloy [4-7].

Many studies in literature are focused on the formation of bulk nanocrystalline HEA/ CCAs with an accompanied enhancement of mechanical properties. In a number of investigations nanocrystalline and often nanotwinned HEA/CCAs are formed through high pressure torsion [8-11]. Nanocrystalline HEA/CCAs formed by severe plastic deformation do not remain single phase on annealing. Formation of additional phases was reported on annealing at 723 K for 5 minutes [8] in contrast to 500 days of annealing of the same alloy with several µm grain size at 773 K and 973 K [12, 13]. Faster formation of secondary phases is attributed to an increase in defect density such as grain boundary and dislocation density due to severe plastic deformation [8], that lead to increase in the number of nucleation sites and provide faster diffusion paths leading to a faster kinetics [14].

Thin film deposition route could be a suitable alternative to severe plastic deformation to study second phase formation due to the reasons listed below. Firstly, nanocrystalline thin films contain a larger area of interfaces that are potential nucleation sites for second phases. Secondly, faster diffusion through interfaces could lead to faster kinetics of phase formation. Thirdly, metastable phases could form due to non-equilibrium process of thin film deposition. Fourthly, thin films on substrates are often under high tensile/ compressive residual stresses – that could influence second phase formation.

The phase diagram, microstructure and thermal stabilities of the bulk quinary (CoCrFeMnNi) and quaternary (CoCrFeNi) HEA/CCAs are studied extensively in literature [2]. However, there are only a few articles on fcc CoCrFeNi and CoCrFeMnNi thin film counterparts [15-20]. Limited data available in the literature suggests that the quaternary CoCrFeNi alloy forms nanocrystalline thin films [17, 18]. Recently, atom probe tomography (APT) samples of quinary alloy with a nanocrystalline microstructure were deposited on a Si micro-tip array [21-23] by magnetron co-sputtering. The nanocrystalline APT tips showed formation of secondary phases at 573 K which is 150 K lower than the previously reported temperature. Given the limited studies on HEA/CCA thin films, there is a huge scope to investigate phases and the microstructure of nanocrystalline thin films made of alloys that form single-phase solid solution in bulk form.

In the present study, CoCrFeNi thin films were deposited on Si/SiO$_2$ and c-sapphire substrates by magnetron co-sputtering. Phases present and microstructure in the as deposited and annealed films are analyzed by X-ray diffraction (XRD) and transmission electron microscopy (TEM). Results are discussed in comparison to the bulk microcrystalline and nanocrystalline HEA/CCA with a similar composition. In addition, agreement with equilibrium phase diagram calculated using different computational approaches is discussed.

## 2. Experimental Procedure

Films were deposited on 100 mm diameter wafers of Si(001)/SiO$_2$ (Siegert GmbH, Aachen, Germany) and on 50 - 100 mm diameter c-sapphire wafers (basal plane (0001) α-Al$_2$O$_3$, one side epipolished, CrysTec GmbH, Berlin, Germany). The Si wafers had 140 nm SiO$_2$ thermal oxide on both sides. The c-sapphire substrates were used as-received without any further wet-chemical surface treatment. Thin films (thickness between 200 nm and 1 μm (Table S1)) were deposited by magnetron co-sputtering of confocally placed 100 mm diameter pure metal targets of Cr (99.95 wt%, MaTecK, Germany), Fe (99.99 wt%, Evochem, Germany), Co (99.99 wt%, MaTecK) and Ni (99.995 wt%, K.J. Lesker, USA). The base pressure of the ultra-high vacuum (HV) system was 2.7 x$10^{-6}$ to 9.3 x $10^{-7}$ Pa and the Ar pressure during sputtering was 0.67 Pa. Both types of substrates were cleaned in the deposition chamber prior to film synthesis with low power radio frequency (RF) Ar ion bombardment with closed shutters. The power to each magnetron sputter source was optimized to obtain an equal atomic concentration of each element in the deposited film. The overall deposition rate was 0.13 nm/s and the sample stage rotation was set to 20 rpm to obtain homogeneous films. Deposition temperatures of 298 K and 573 K were selected.

Deposited films were characterized for phase analysis in a Seifert diffractometer equipped with an ID3003 generator, a poly-capillary beam optic, a 2-circle goniometer, using Co-K$\alpha$ radiation and an energy dispersive point detector. The θ-2θ overview measurements were carried out with a step size $\Delta 2\theta$ of 0.03° and a count time of 10s/step. To avoid signal from the substrate an offset of 10° was applied to the θ axis.

The film composition was characterized by energy dispersive X-ray spectroscopy (EDS) in a scanning electron microscope (SEM). Further, the microstructure was imaged in a high-resolution field emission gun Scanning Electron Microscope (FEG-SEM) Zeiss Gemini 500. The TEM samples were prepared by the Ga focused ion beam (FIB) lift-out method (Scios2, Thermo Scientific). The coarse milling was done with an accelerating voltage of 30 kV followed by final polishing at 5 and 2 kV. An aberration corrected TEM (Titan Themis 60-300, Thermo Fisher Scientific) was used for imaging, diffraction and EDS analysis. The electron beam was accelerated to 300 kV. Surface roughness of a few as deposited films was measured using a JPK Bio atomic force microscope (AFM) (Bruker Nano GmbH, Berlin, Germany) in tapping mode.

Residual stress measurements were carried out using the $\sin^2\psi$ method [24]. These measurements were made with a Bruker D8 GADDS diffractometer equipped with a poly-capillary beam optic, a 4-circle goniometer, using Co-K$\alpha$ radiation and an area detector. The range of ψ values spanned from -70 to 70° with a total of 15 steps for the $\{111\}_{fcc}$ reflex. X-ray elastic constants for γ- Fe were taken from the internal material database of the Bruker Software LEPTOS.

Residual stresses were also calculated using the wafer curvature technique [25, 26]. These measurements were carried out on 10 mm x 10 mm thin film samples and the mean value of more than 10 measurements was calculated. Average of curvatures of uncoated substrates with similar sizes were subtracted from these values to obtain the net curvature due to thin film stresses alone. The residual stress in the thin film was calculated using the standard Stoney equation [25, 27].

Annealing treatments were carried out in a custom-built vacuum chamber with a pyrolytic BN heater under a vacuum of $10^{-1}$ - $10^{-3}$ Pa at temperatures below 1273 K. The chromium oxide scale on the surface of the annealed samples was removed by Ar ion surface polishing in a Gatan Precision Etching and Coating System (PECS) in the following conditions: acceleration voltage of 5 kV, 210 µA milling current, 10° incidence angle of the ion beam for ~ 5 to 20 minutes before SEM and EBSD investigations.

## 3. Results

### 3.1. The fcc phase and an additional phase peaks in XRD patterns of as-deposited films

The as-deposited films displayed fcc peaks [28] under θ-2θ measurements (Fig. 1a). Due to a strong {111} fiber texture, the {111} peak in XRD was intense, other peaks were weaker but still observed. Analysis of the XRD θ-2θ scans provides an fcc lattice parameter value of 0.355 nm (calculated without standards) which is consistent with that reported for the bulk quaternary alloy [5]. In addition to the peaks of the fcc phase, another peak (2θ value of ~48.5°) was observed corresponding to a d-spacing of ~ 0.215 nm in most of the as-deposited samples (Fig. 1a). The intensity ratio of this additional peak to the $\{111\}_{fcc}$ peak was less than 10% (Table S1). This peak was also accompanied by a broad peak between $\{111\}_{fcc}$ and $\{200\}_{fcc}$ peaks (angles of ~ 51.5 – 60.5°; d-spacing between 0.21 and 0.18 nm). Although maxima other than the one around ~ 48.5° could not be identified clearly as peaks, their presence was evident in most of the as-deposited films (Fig. 1b, a total of 14 films were investigated). To help identifying the possible phase leading to these broad reflections, we tried matching observed peaks with possible peaks in binary phase diagrams between each pair of elements in the HEA as well as the reported phases in the bulk equiatomic CoCrFeNi HEA. The peaks clearly did not match with the $Cr_2O_3$ peaks[29], the superlattice reflection peaks from intermetallics in the system Ni-Fe [30], hcp Co [31] or the bcc phase [31] observed in CoCrFeNi alloy systems (Fig. 1d). We confirmed that the peak was not related to the substrate (Fig. 1c). Moreover, since no reaction with Si was observed in TEM (Fig. 2a and 2e) we exclude that any additional peak is related to Si compounds. In addition, these additional peaks were also present in 1 µm thick films on c-sapphire substrates (Fig. 1a,b; For details on Film No. see Table S1) leading to the exclusion of peaks from reaction products of the alloy with the substrate.

### 3.2. Additional peaks in the TEM-SADP analysis of as deposited films

Additional peaks were also observed in selected area diffraction pattern (SADP) acquired in the TEM (Fig. 2c). To precisely measure the reciprocal spacing of the reflections, peaks from the Si substrate were used for the calibration (Fig. 2c). Radial integration of the intensity of SADPs from the film (without Si peaks) clearly confirms an additional peak at a d-spacing of 0.215 nm, consistent with XRD (marked by vertical line in Fig. 2d) within the expected precision. The peak in XRD at 0.19 nm is seen as the broad base of the $\{111\}_{fcc}$ peak (also marked as a vertical line in Fig. 2d). Hence observations of XRD and TEM listed above are consistent with peaks from an additional phase present in small fraction along with the fcc phase in the thin film. In the literature of CrMnFeCoNi ('Cantor') alloy [14, 32, 33], a large number of peaks close to $\{111\}_{fcc}$ and $\{200\}_{fcc}$ correspond to the Cr-rich intermetallic σ phase [34-36]. XRD profiles (Figs 1a, b) match very well with σ phase peaks. The peak at 0.215 nm corresponds to the $(321)_\sigma$ and the broad peak at 0.19 nm is a convolution of several σ peaks that occur between $\{111\}_{fcc}$ and $\{002\}_{fcc}$.

### 3.3. Microstructure and composition homogeneity of as deposited films

The SEM-EDS showed a uniform composition within ± 1 at % of all elements compared to the nominal composition (Table S1). In addition, the TEM-EDS revealed no signs of segregation (Fig. 2e). The average roughness of the films measured by AFM was lower than 5 nm.

Cross sections of films (Fig. 2a) displayed columnar grains, where grain diameter parallel to the film surface is much smaller compared to the column height which is of the order of film thickness. The average diameter of the columnar grains measured using cross sectional TEM image (Fig. 2a) is smaller than 100 nm for as-deposited films, whereas the film thickness and the height of columnar grains varied between 200 nm and 1 μm. The grain diameters of as-deposited films were measured using the contrast due to the depressions near grain boundaries at the film surface in the SEM secondary electron images (see Figs. 3a and b, Table S1). A high density of twin boundaries and stacking faults was observed in cross section (Fig. 2b). Multiple closely located σ phase peaks near the $\{111\}_{fcc}$ and $\{200\}_{fcc}$ peaks lead to inability in separation of diffraction spots from the σ phase alone using SADP apertures.

The grain size of the σ phase is expected to be in nanometers due to a nanocrystalline structure of fcc phase, leading to difficulty in spatially resolving the σ phase in conventional TEM images. However, the hypothesis of the presence of an additional σ phase is also confirmed indirectly by annealing as explained in section 3.5.

We also exclude the additional peaks ascribed to the σ phase to be due to a high density of stacking faults leading to hcp reflections [37]. There was no one-to-one correspondence between presence of a high density of stacking faults, twin boundaries and appearance of these additional peaks. For example, 500 nm thin films on c-sapphire that contained extensive twin boundaries and stacking faults both in as deposited and annealed conditions showed only the fcc phase (film no. 12 and 13 in Table S1 and Fig. 1b). Also, these additional peaks (ascribed to the σ phase) were only observed in films without an oxide layer and disappeared in annealed films with an oxide layer but still containing twin boundaries/ stacking faults (Fig. 4b).

### 3.4. Residual stresses in as deposited films

Residual stresses in as deposited and annealed films on Si/SiO$_2$ measured by the wafer curvature technique and sin$^2$ψ technique are listed in Table S2. Both techniques lead to high residual stresses between 700 MPa and 1.2 GPa in all as-deposited films. Annealing at 773 K results in further increase in the residual stress that can reach values above 1.2 GPa. A large scatter in individual measurements result in a considerable error in these wafer curvature values (Table S2) [25, 27]. Within the error limits, values from wafer curvature technique agree with measurements using XRD (Table S2).

### 3.5. Disappearance of additional phase peaks in annealed films

Typical XRD patterns of the films annealed at temperatures between 573 and 1073 K are shown in Fig. 4a. The fcc phase remained stable on annealing up to a temperature of 1223 K for films on Si/SiO$_2$ and up to at least 1373 K for films on c-sapphire. Annealing under low vacuum (10$^{-1}$ Pa at the annealing temperature) above 623 K or medium vacuum (10$^{-3}$ Pa at the annealing temperature) above 673 K leads to the formation of a thick Cr$_2$O$_3$ scale on the alloy surface and a few oxide inclusions inside the film. XRD peaks from the rhombohedral Cr$_2$O$_3$ phase were visible at d-spacings of 0.267 and 0.167 nm in the annealed film (Fig. 4a). Interestingly, formation of Cr$_2$O$_3$ always resulted in the decrease in intensity and sometimes the disappearance of the broad XRD peaks from the σ phase. This was further confirmed by comparing of the same film annealed at two different levels of vacuum at 673 K (10$^{-3}$ and 10$^-$

$^1$ Pa; Fig. 4b). This observation suggests that the additional peaks that consistently occur in most of the as-deposited films in the present study were from a Cr-rich phase.

## 4. Discussion
### 4.1. On the occurrence of the σ phase at RT

In the present study, occurrence of a small but a definite fraction of the σ phase was evident in most of the as-deposited films on Si/SiO$_2$, while a single-phase fcc was only observed in less than 700 nm thin films on c-sapphire. A detailed summary of other studies on nanocrystalline alloy where σ phase was reported is given in Table S3. In addition, formation of Cr$_2$O$_3$ resulted in the decrease in intensity or sometimes the disappearance of the σ phase peaks. The σ phase is a topologically close packed phase [34, 35] with body centered tetragonal crystal structure showing a set of closely spaced Co-Kα diffraction peaks between d-spacings of 0.22 – 0.17 nm (2θ of 48 and 60°) (see Fig. 1d). XRD data in the present study suggests a crystal structure with a=b= 0.882 nm and c= 0.458 nm (similar to the crystal structure of binary Cr$_3$Ni$_2$). Lattice parameters are consistent with various studies in literature (tetragonal P4$_2$/mnm, a ~ 0.88 nm and c ~ 0.44 – 0.46 nm) [11-14, 38, 39] (Table S3).

The σ phase is reported very often in experimental studies on CoCrFeMnNi annealed at temperatures between 773 and 1173 K [8, 12-14, 38]. Experimental studies on CoCrFeNi [40] reveal a single-phase fcc solid solution at higher temperatures. To our knowledge, this study is the first experimental evidence for the σ phase in equiatomic CoCrFeNi alloy at room temperature although Cr rich compositions are reported to show the σ phase [41]. Most of the studies on CoCrFeNi alloy thin films [17],[18], [20] report a single-phase fcc solid solution. However, it should be noted that non-equilibrium phases could be present in thin films – although not in the bulk due to differences in the processing route (e.g. high condensation rate for thin film deposition). For example, FeCr binary alloy thin film was completely made of the nanocrystalline σ phase near equiatomic composition even at RT [42], although the σ phase is a high temperature phase according to the equilibrium phase diagram.

It is notable that the calculated phase diagrams predict the σ phase even at lower temperatures: for example at all temperatures below 913 K in CoCrFeNi [43], and below 873 K [38, 40] in CoCrFeMnNi alloys. In an extensive computational study on the quinary space using CALPHAD approach and the TCHEA1 database [40], a multiphase region of fcc + bcc 1 +

bcc 2 + σ phase was predicted at temperatures less than 873 K, even for the equiatomic CoCrFeMnNi "Cantor" alloy. Here, bcc 1 and bcc 2 are bcc disordered phases with different lattice parameters and compositions. The bcc1 phase starts to appear below 773 K and the σ phase below 873 K. Similarly, the quaternary CoCrFeNi quaternary alloy shows a single phase solid solution only above ~ 913 K, where additional bcc 1 phase forms below 913 K, fcc 2 phase below ~ 850 K and bcc 2 phase below ~ 690 K. The simulation results in that study [40] were only reported at temperatures above 600 K for the CoCrFeNi alloy studied in this paper. At 600 K, the alloy is expected to contain fcc, bcc1 and fcc 2 phases according to the calculated phase diagrams. However, the isopleth between CoCrFeNi and Mn, σ phase is present at 600 K at Mn concentration between 2.5 % and 92.5%. The σ phase field expands at decreasing Mn concentrations with decreasing temperature – indicating that the σ phase could be stable in CoCrFeNi alloy at 300 K. Furthermore, the study [40] clearly mentions the underestimation of the σ phase field by CALPHAD calculations compared to experimental data [40]. Above considerations naturally lead to a possibility of the presence of the fcc + σ phase at RT at equilibrium in CoCrFeNi alloy. The reason for the absence of the bcc phase at RT is given by another study [38] on the quinary alloy as explained below.

In reference [38], ThermoCalc database TCFE2000 was used to predict phases in CrMnFeCoNi alloy. Calculations showed fcc + bcc 1+ bcc 2 phase at less than 973 K when database for all phases was considered. But, when bcc phase was suspended from the system, the alloy forms a fcc + σ two phase region below 913 K down to RT. Hence, it was inferred that the bcc phase is the stable phase, σ phase despite being quite stable for long hours of annealing is still considered as a metastable phase.

Similarly in an early study using CALPHAD database – where information on phases was extrapolated using binary systems, fcc + bcc 1 + bcc 2 phase were observed for several alloy compositions that are reported to show the σ phase under experimental conditions [44].

From the simulation results discussed above as well as the data on thin films in the present study, it is clear that the σ phase appears more frequently under experimental conditions compared to computational studies – in many cases replacing the bcc phase reported by equilibrium phase diagram calculations. The reasons for the same are explained in the subsections below.

### 4.2. Factors influencing formation of the σ phase

The σ phase could appear in thin films due to several reasons. Firstly, the film has a nanocrystalline microstructure (Sec. 3.3). Secondly, films in this study are under large tensile residual stresses at room temperature (Sec 3.4). Thirdly, deposition was carried out from pure elemental sources and segregation of individual elements at the atomic level due to the deposition process cannot be excluded. The contribution of each of these factors is discussed in detail below.

#### 4.2.1. Nanocrystalline, nanotwinned microstructure

In most of the literature studies on the bulk CoCrFeNi alloy, the σ phase is observed at triple junctions [14], grain boundaries [8, 12, 13], incoherent and coherent twin boundaries (annealing twins) [14] and in the fcc matrix – often connected to a dislocation network [45, 46]. Experimental data in [38] clearly showed formation of the σ phase at defects such as triple junctions, grain boundaries, incoherent twin boundaries and dislocations. Combined with the computational results mentioned above, it is clear that the kinetics for the formation of the metastable σ phase becomes favorable in the presence of defects. Similarly, although it was not noted, the σ phase precipitates in reference [40] also appear mostly at grain boundaries especially in alloys with a low volume fraction of sigma phase. Hence, the discrepancy between experiments and thermodynamic calculations in terms of an enhanced σ phase field is certainly related to the heterogenous nucleation at defects combined with the corresponding enhanced rate of diffusion – factors that are not accounted for in calculated phase diagrams.

In the present study, sputter deposition on substrates at room temperature and 573 K lead to the formation of σ phase due to the large area of interfaces in the microstructure. A high density of columnar grain boundaries, incoherent and coherent twin boundaries within the columns (growth twins) form potential sites for nucleation of the σ phase during thin film deposition. The present study suggests that the σ phase forms even at room temperature provided there are sufficient nucleation sites and faster diffusion paths (interfaces) available, both factors leading to the rapid kinetics of formation [8, 14]. However, formation of bcc phase that is expected to be stable at RT according to the calculations, is much slower than the σ phase formation [38]. Note that bcc phase precipitates formed homogenously in the matrix in reference [40] (in compositions where they appear). Hence the formation of bcc precipitates requires lattice diffusion, which is too sluggish at RT. Hence, bcc phase formation does not take place under experimental conditions used in the presence study.

### 4.2.2. Tensile residual stresses

Films in the present study are under huge tensile stresses. Residual stresses at room temperature are expected to be tensile on cooling from the deposition temperature of 573 K, due to the stronger contraction of the film compared to the substrate because of a larger thermal expansion coefficient of the film. Tensile stresses can also originate from the coalescence of islands during the film growth [47]. Absence of tensile stress relaxation after deposition could be attributed to a slow surface and grain boundary diffusion in the alloy [48] at room temperature. This is evident from the microstructure showing columnar grains where grain diameters (between 20 and 100 nm) are an order of magnitude less compared the film thickness (200 nm to 1 µm). Such microstructure is similar to the high melting metal alloy thin films with a low adatom mobility [49]. In contrast, low melting metal alloy thin films with high adatom mobility show considerable grain growth at the deposition temperature leading to grain diameters up to 2 to 3 times the film thickness [49].

In addition, segregation of Cr at the atomic level (although not observed down to a few nanometers, see Fig. 2e) could significantly alter residual stresses and grain size of the thin film. Literature on alloy thin films indicates the influence of composition [50] on the grain size, magnitude and sign of the residual stress developed. Grain boundary segregation [50], chemical ordering [51] and phase separation [52] are reported to manipulate the thin film stress states. Conversely, residual stresses due to coherency strains could lead to phase transformation such as observed in the formation of bcc Zr in a Nb/Zr multilayers at lower thicknesses [53]. However, studies addressing the influence of residual stress on phase separation in polycrystalline thin films with incoherent interfaces are less addressed in literature.

In the present study we note that the volume of the σ phase per atom (considering a = 0.882 nm and c = 0.458 nm) at room temperature is 0.0117 $nm^3$, whereas the same parameter for the fcc phase is 0.0112 $nm^3$. These simple calculations are consistent with the tensile residual stress favoring the precipitation of the bulkier σ phase. Testing of this line of reasoning requires systematic changes to the deposition process to alter these residual stresses. In addition, phase analysis of the annealed films – freestanding (without residual stresses) and supported on the substrate (with residual stresses) are likely to reveal the role of residual stresses. These issues will be addressed in a future study.

### 4.2.3. Influence of local composition

It should be acknowledged that the σ phase is present in most of the deposited films but not in all films (see Table S1). Films with nominally the same composition (composition difference of elements < 1 at %) deposited under the same deposition conditions (same base vacuum, sputtering pressure, substrate temperature, power applied to sputtering targets and the film thickness) show differences in terms of the presence / absence of the σ phase (Table S1). As mentioned in the last section, local chemical deviations have been observed in alloy thin films in literature [48 - 50]. It could be speculated that the segregation of Cr at the atomic level could lead to these differences, which will be further investigated by APT in the future work. From the Sections 4.2.1 – 4.2.3, it is clear that (i) nanocrystalline microstructure with a large area of interfaces, (ii) tensile residual stresses and (iii) possible local excess of Cr due to the deposition from individual pure metal sources all favor the formation of a Cr-rich and bulkier σ phase in the otherwise predominantly fcc CoCrFeNi alloy.

### 4.3. Disappearance of the σ phase on formation of $Cr_2O_3$

As thin film deposition by sputtering at room temperature is a non-equilibrium process, experiments are not self-sufficient to prove that the σ phase observed is a metastable phase or an equilibrium phase at room temperature. However, the formed σ phase remains stable on annealing up to 673 K (for up to 72 hours) in the absence of oxidation. Presence of the σ phase above 673 K could not be confirmed independent of oxidation. Studies are ongoing to anneal films at higher temperatures without oxidation. The thickness of the $Cr_2O_3$ layer for films annealed under vacuum of $10^{-3}$ Pa at 973 K and 1223 K for 2 hours was 90 nm and 100 nm, respectively. Formation of $Cr_2O_3$ led to a slight depletion of Cr in the matrix. A typical composition of an annealed film (all in at %) measured by TEM-EDS was 23 Cr, 27 Fe, 25 Co and 25 Ni. Disappearance of the σ phase on consumption of Cr by $Cr_2O_3$ formation suggests that a high Cr content in the solid solution is required for the formation of the σ phase, being consistent with the literature [12, 14]. We also note that as-deposited films with less than 24 at. % Cr (Film Nos. 4, 9 and 11 in Table S1) contain the σ phase, whereas an annealed film with 23 at. % Cr does not show any σ phase, implying that nucleation events and local chemical fluctuations (segregation or short range ordering) during deposition rather than the absolute composition alone play the decisive role in the appearance of the σ phase in as-deposited films.

## 5. Summary and conclusions

In summary, nanocrystalline CoCrFeNi thin films were successfully deposited on Si/SiO$_2$ and c-sapphire substrates. In addition to the fcc phase, most of the films contained an additional σ phase, the strongest peak intensity which was less than 10% of the $\{111\}_{fcc}$ peak. The XRD peaks corresponding to this additional phase disappeared when a Cr$_2$O$_3$ layer forms upon annealing but remained in the absence oxidation. Observations clearly support the hypothesis that the additional phase is a Cr-rich phase and that a high concentration of Cr is required for the formation of the σ phase, consistent with the bulk HEA/CCA literature. In contrast to bulk CoCrFeMnNi [8, 12], this additional phase is formed even at room temperature and in the as-deposited films due to a large number of nucleation sites (such as a dense network of grain boundaries, triple junctions and twin boundaries) and a faster formation kinetics promoted by the finer microstructure. In addition, high tensile residual stresses inherently present in these thin films and possible composition inhomogeneity at the atomic scale due to deposition from individual pure elemental sources possibly trigger the formation of the σ phase at room temperature. The observation of the σ phase in the equiatomic CoCrFeNi alloy thin films at RT contrasts with the formation of the same phase at temperatures above 773 K in bulk nanocrystalline HEA [8]. However, results are consistent with data predicted by equilibrium phase diagrams [39, 40].

**Acknowledgements:**

The study was carried out under the DFG-ANR collaborative project ''Analysis of the stability of High Entropy Alloys by Dewetting of thin films'' under the grant LU1175/22-1 (RU Bochum), DE796/11-1 (MPIE) and ANR-AHEAD-16-CE92-0015-01 (CINaM and CEMEF). The authors gratefully acknowledge AFM measurements by X. Fang. Critical discussions with Prof. Paul Wynblatt (Carnegie Mellon University, Pittsburgh) are gratefully acknowledged.

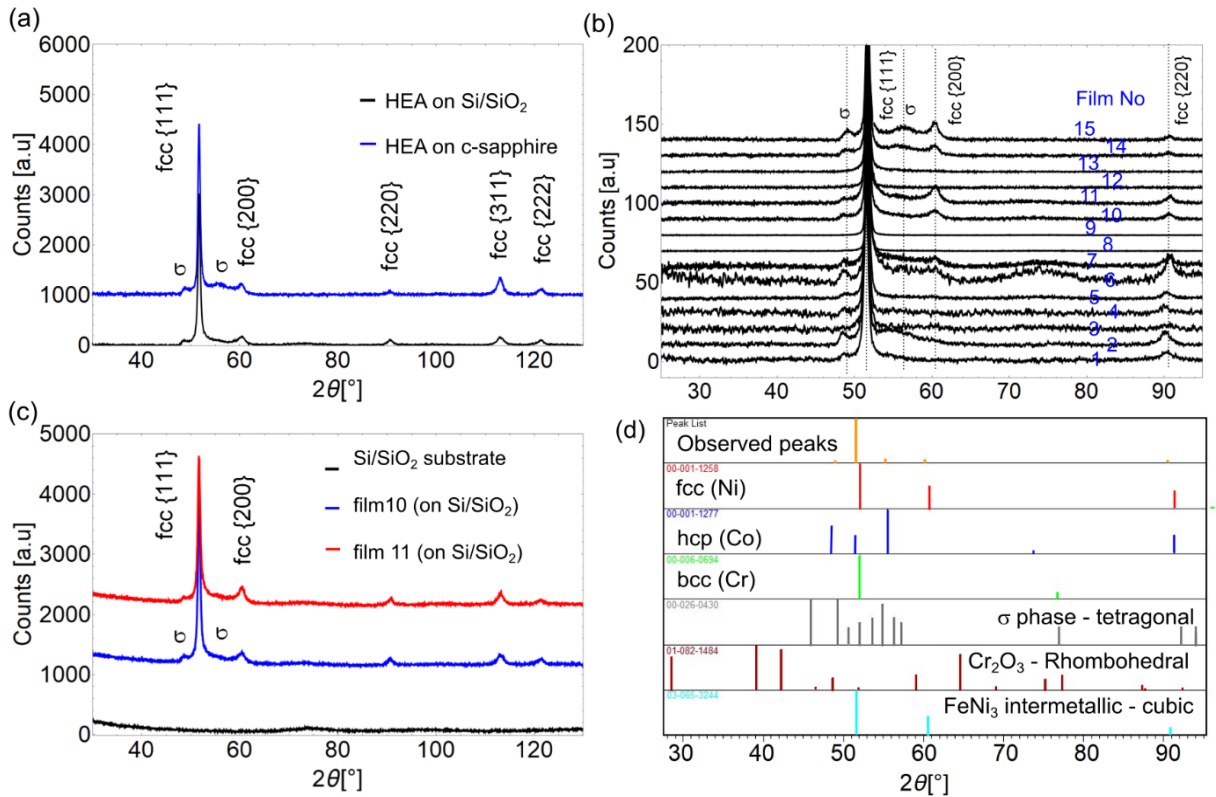

Fig. 1: CoCrFeNi thin films on Si/SiO$_2$ and c-sapphire (a) typical XRD patterns (θ-2θ scans) of films on Si/SiO$_2$ and c-sapphire showing σ peaks in addition to fcc peaks and (b) XRD θ-2θ scans from all deposited films (for details on the Film No. see Table S1), most of the films show σ phase (c) XRD θ-2θ patterns from films on Si/SiO$_2$ and a bare Si/SiO$_2$ substrate. (d) Reference line patterns for the possible phases in the quaternary alloy. The literature for reference patterns in Fig.1d are taken from – bcc Cr [31], hcp Co [28], fcc Ni [28], tetragonal σ phase [34-36], Cr$_2$O$_3$ (rhombohedral) [29] and FeNi$_3$ intermetallic [30].

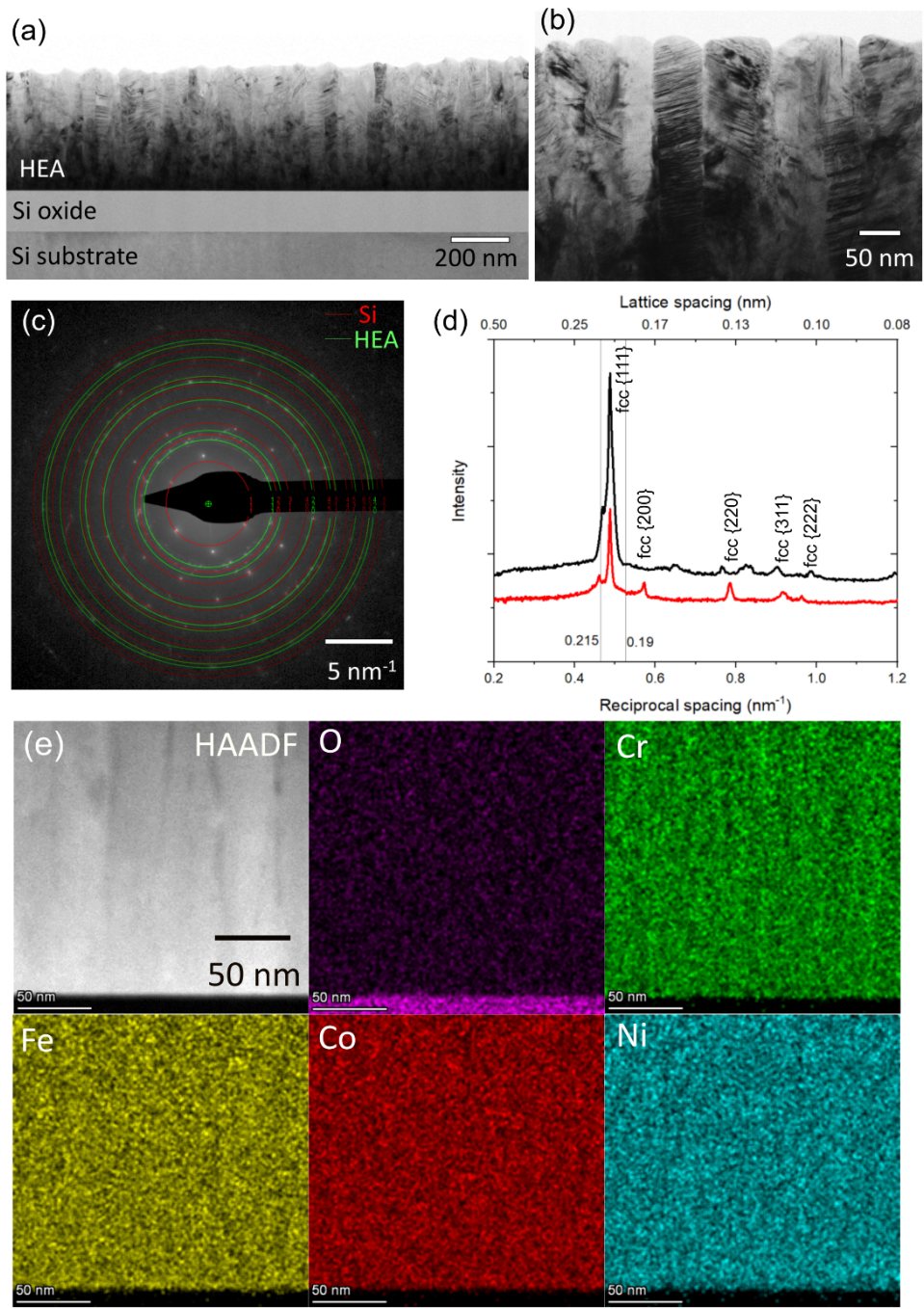

Fig. 2: (a) Cross section bright field image, (b) a magnified image showing columnar grains densely populated with coherent twin boundaries and stacking faults. (c) TEM selected area diffraction pattern (SADP) and (d) corresponding radial integration of SADPs (Black and red curves are the measurements in a film on Si/SiO$_2$ in two different locations) (e) HAADF STEM and EDS maps of an as deposited film on Si/SiO$_2$ substrate show a homogenous composition.

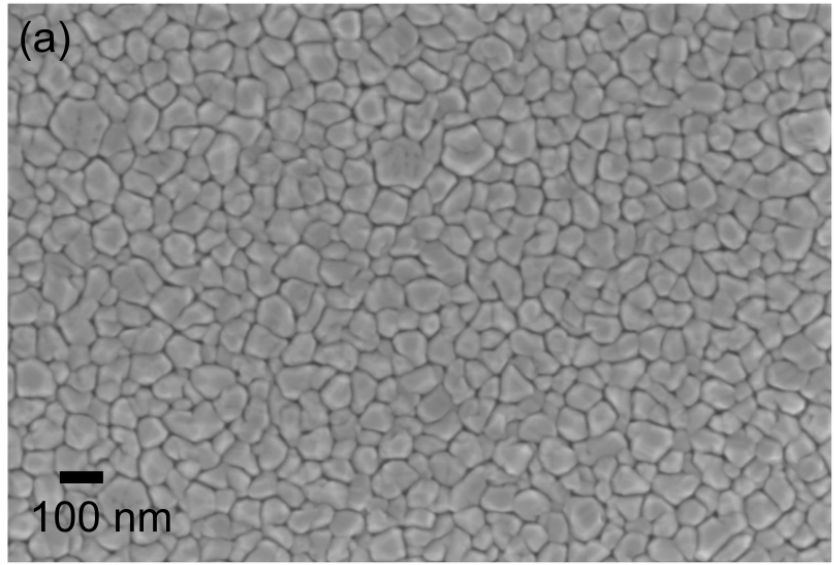

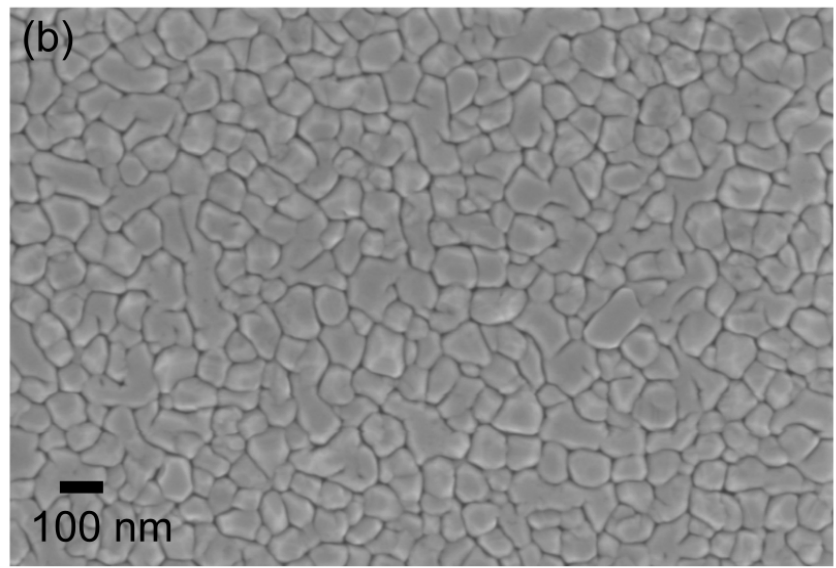

Fig 3: Typical plan view SEM images of the films deposited on (a) Si/SiO$_2$ and (b) c-sapphire at 573 K.

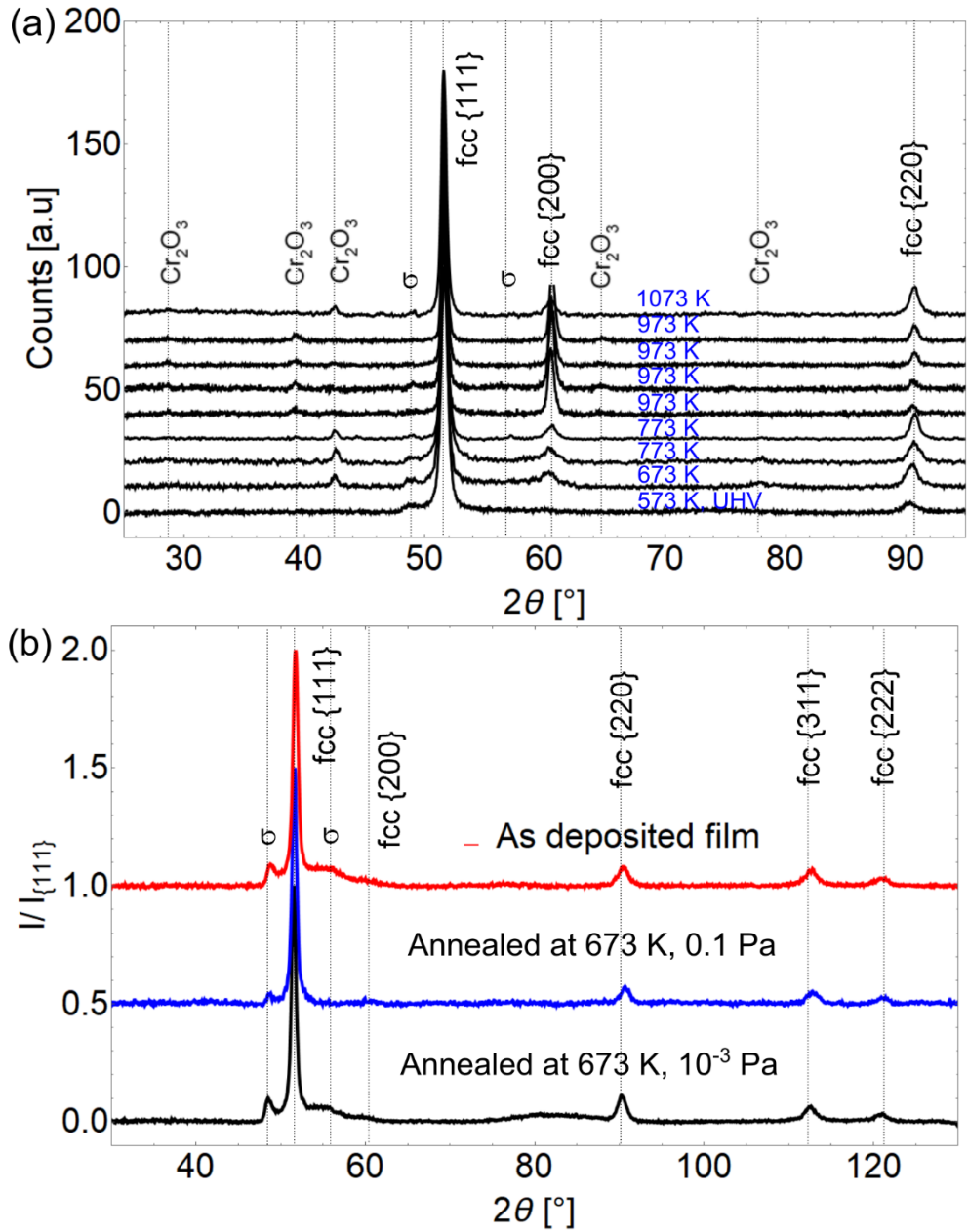

Fig. 4: Formation of $Cr_2O_3$ leading to the disappearance of the σ phase (a) XRD on films annealed at different temperatures above 573 K and (b) XRD on film annealed at 673 K at different levels of vacuum.